\documentstyle[12pt, epsfig]{article}
\newcommand{\be}{\begin{equation}}
\newcommand{\ee}{\end{equation}}
\newcommand{\bea}{\begin{eqnarray}}
\newcommand{\eea}{\end{eqnarray}}
\newcommand{\bd}{\begin{displaymath}}
\newcommand{\ed}{\end{displaymath}}

\begin{document}
\bibliographystyle{physics}
\renewcommand{\thefootnote}{\fnsymbol{footnote}}

\begin{titlepage}

\begin{center}

{\large{\bf Matrix Elements of Four-quark Operators Relevant to 
Lifetime Difference $\Delta \Gamma _{B_s}$ from QCD Sum Rules}}

\vspace{1.2cm}

C.-S. Huang$^a$, Ailin Zhang$^a$ and Shi-Lin Zhu$^{a,b}$

\vspace{0.8cm}
$^a$ Institute of Theoretical Physics, Academia Sinica, P. O. Box 2735,
Beijing 100080, China\\
$^b$ Department of Physics, University of Connecticut,
Storrs, CT 06269 USA

\end{center}

\vspace{1.0cm}

\begin{abstract}

We extract the matrix elements of four quark operators $O_{L,S}$ relevant
to the $B_s$ and ${\bar B}_s$ life time difference from QCD sum
rules. We find the vacuum saturation approximation works reasonably well,
i.e., within $10\%$. We discuss the implications of our results and
compare them with the recent lattice QCD determination.

\vskip 0.5 true cm

PACS Indices:  12.38.Lg, 12.39.Hg, 13.25.Hw, 13.30.-a.\par

\end{abstract}

\vspace{2cm}
\vfill
\end{titlepage}

\section{Introduction}
The recent results on CP violation in 
$B_d$ - $\bar B_d$ mixing have been reported by the BaBar and Belle
Collaborations 
\cite{Osaka} in the ICHEP2000 Conference. More experiments on B physics
have been planned
in  the present and future B factories \cite{bf}. Theoretical efforts to
improve predictions
and reduce uncertainties are expected and needed.
It is well-known that mixing in neutral B meson systems provides a good
place to examine CP violation
as well as flavor physics in the standard model and beyond. For example,
the mass difference between 
the mass eigenstates of neutral $B_d$ meson, $\Delta M_{B_d}$, gives an
important constraint on CKM
matrix element $V_{td}$ and the first indication of large mass of top
quark. Similarly, 
the mass difference between the mass eigenstates of neutral $B_s$ meson,
$\Delta M_{B_s}$, which will
be precisely measured in the near future would give an valuable constraint
on CKM
matrix element $V_{ts}$. The another important observable for mixing in
neutral B meson systems is
the lifetime difference between  the mass eigenstates of neutral B mesons,
$\Delta \Gamma_{B_d}$
or $\Delta \Gamma_{B_s}$.  The ratio $|V_{ts}/V_{td}|^2$ can be extracted
from the measurement of
$\Delta \Gamma_{B_s}$ \cite{bp}. The width difference of $B_d$ mesons is
CKM suppressed and consequently
not easy to be observed. In contrast, for $B_s$ mesons the width
difference is large enough to
be measured \cite{bi} and has been recently measured \cite{wa} with low
statistics. Hopefully, it will
be measured with high statistics in the near future.

As usual, The light $B_s^L$ and heavy $B_s^H$ mass eigenstates are defined
by
$$|{B^{L,H}_s}\rangle=p|{B_s^0}\rangle\pm q|{\bar B_s^0}\rangle,$$
where $|{B_s^0}\rangle$ and $|{\bar B_s^0}\rangle$ are the flavor
eigenstates. 
The mass difference and the width 
difference between the physical states are given by
$$\Delta m\equiv M_H-M_L, \Delta\Gamma\equiv\Gamma_H-\Gamma_L.$$
Because $|\Gamma_{12}|\ll|M_{12}|$ for $B_s$ mesons \cite{beneke}, 
to the leading order in 
$|\Gamma_{12}/M_{12}|$, 
$\Delta m_B=2|M_{12}|,
\Delta\Gamma_B=\ 2\Re(M_{12}\Gamma_{12}^*)/|M_{12}|.$
Neglecting very small CP violating corrections, the width difference for
$B_s$ mesons in SM has been
given  
\cite{beneke,ben}
\begin{equation}\label{dgabc}
\left(\frac{\Delta\Gamma}{\Gamma}\right)_{B_s}=
\left(\frac{f_{B_s}}{210~{\rm MeV}}\right)^2
\left[0.006\, B(m_b)+ 0.150\, B_S(m_b) - 0.063\right], 
\end{equation}
where $f_{B_s}$ is the decay constant of $B_s$, B and $B_S$ are the bag
parameters related to
 the four quark operators
$O_L$ and $O_S$ (see below). These hadronic quantities need to be
calculated by non perturbative methods such as lattice, QCD sum 
rules, Bethe-Salpeter approach, etc. 

The similar quantities related to $B^0_d-\bar B^0_d$ mixing have been
estimated by Narison {\it et al}
within the traditional QCD sum rules approach\cite{narison} at
$\alpha_s$ order. Their conclusion is that the vacuum saturation
values $B_B\simeq B_{B^*}\simeq 1$ are satisfied within $15\%$.
Their sum rules are constructed through two-point correlation functions
and depend on some phenomenological assumptions. In this letter we shall
calculate the matrix elements of four-quark operators relevant to the
$B_s$ meson lifetime difference through QCD sum rules in HQET. The sum
rules are constructed with three-point correlation functions.
Our calculation is carried out at the leading order in $1/m_b$ expansion
in HQET for simplicity. In ref.\cite{narison} the effects of condensates
are absorbed like other factorizable corrections into the contribution to
$f_B$ and the available result of \cite{pich}(though not
explictily said) has been used as the effects are small compared to the
perturbative corrections. In our sum rules the nonperturbative
contributions of condensates are explicitily included and the numerical
results confirm the smallness of these corrections (see below).  
\section{Theoretical formalism}

We employ the following three-point Green's function,
\begin{equation}\label{cor}
\Gamma^O(\omega, \omega')=i^2\int dxdye^{ik'\cdot x+ik\cdot y}
\langle0|{\cal T}[\bar{s}(x)\gamma_5 h_v^{(b)}(x)]O_{L,S}(0)
[\bar{s}(y) \gamma_5h_v^{(b)}(y)]|0\rangle~,
\end{equation}
where $\omega=v\cdot k$, $\omega'=v\cdot k'$; $h_v^{(b)}$ is the b-quark
field in the HQET with velocity $v$.  And $O_{L,S}$ denotes the
color-singlet
four quark operators. They are 
\begin{equation}
O_L=\bar{b}\gamma^\mu (1-\gamma_5) s\bar{b}\gamma_\mu (1-\gamma_5) s,
\end{equation}
\begin{equation}
O_S=\bar{b}(1-\gamma_5) s\bar{b}(1-\gamma_5) s,
\end{equation}
In terms of the hadronic expression, the correlator in
Eq. (\ref{cor}) reads

\begin{equation}\label{pole}
\Gamma^O(\omega, \omega')={F_{B_s}^2\over 4}\frac{\langle {\bar
B}_s|Q_{L,S}|B_s\rangle}
{(\bar{\Lambda}-\omega)(\bar{\Lambda}-\omega')}+ {\rm resonances}~,\\[3mm]
\end{equation}
where $\bar{\Lambda}=m_B-m_b$ and $F_{B_s}$ is the $B_s$ decay constant in
the 
leading order of heavy quark expansion defined as \footnote{Note that
$f_{B_s}$ in Eq. (1) is 
defined by
\be
\label{two}
\langle 0~|~ {\bar s} ~\gamma^\mu~ \gamma_5 ~b~| B^0_s \rangle 
= -i f_{B_s} p^\mu \\
\ee}

\be
\langle 0|\bar{s}(0) \gamma_5h_v^{(b)}(0)|B_s\rangle=-i\sqrt{m_Q}F_{B_s}
\ee

In order to eliminate the contribution from the non-diagonal single pole
terms
and suppress the continuum contribution in Eq. (5), we make double 
Borel transformation to the correlator. The transformation is defined as
\begin{equation}
\label{12}
\hat{B} = \lim_{\begin{array}{c}
-\omega\to\infty\\n\to\infty\\
\tilde{\tau}\equiv\frac{-\omega}{n}~ {\rm fixed} 
\end{array}}
\lim_{\begin{array}{c}-\omega'\to\infty\\m\to\infty\\ 
\tilde{\tau}'\equiv\frac{-\omega'}{m}~ {\rm fixed}
\end{array}}
\frac{(-\omega )^{n+1}}{n!}\left(\frac{d}{d\omega }\right)^n
\frac{(-\omega')^{m+1}}{m!}\left(\frac{d}{d\omega'}\right)^m \; .
\end{equation}
There are two Borel parameters $\tilde{\tau}$ and $\tilde{\tau}'$, which 
appear symmetrically, so $\tilde{\tau}=\tilde{\tau}'=2T$ are taken in the 
following analysis. 

On the other hand the correlator can be calculated at the quark gluon
level. For
example for $O_L$ we may rewrite the right hand side of Eq. (\ref{cor}) as
 
\bea \label{qq}\nonumber
-2\int dxdye^{ik'\cdot x+ik\cdot y}\{ -\mbox{Tr}[\gamma_5 \cdot
iS^{mi}_{b}(x)
\cdot \gamma^\mu (1-\gamma_5)\cdot iS^{in}_s(-y)\cdot \gamma_5
iS^{nj}_{b}(y)
\cdot \gamma_\mu (1-\gamma_5)\cdot iS^{jm}_s(-x)]\\ 
+\mbox{Tr}[iS^{im}_s(-x)]\cdot\gamma_5 \cdot iS^{mi}_{b}(x)\cdot\gamma^\mu
(1-\gamma_5)]
\mbox{Tr}[iS^{jn}_s(-y)]\cdot\gamma_5 \cdot iS^{nj}_{b}(y)\cdot\gamma_\mu
(1-\gamma_5)]
\}
\eea
where $iS^{jn}_s(x)$ is the full strange quark propagator with both
perturbative term
and condensates, $i, j$ etc is the color index. $iS^{nj}_{b}(x)$ is the
leading
order heavy quark propagator which has very simple form in coordinate
space:
\be
iS^{ij}_{b}(x) =\delta^{ij}\int_0^\infty dt \delta (x-vt)
\ee
Note the structure of color flow is quite different for the two terms in
Eq. (\ref{qq}).
For the perturbative part the first and second term is proportional to
$N_c$ and $N_c^2$,
respectively, where $N_c=3$ is the QCD color number. In the limit of
$N_c\to\infty$, the
second term dominates! As shown below, the non-factorizable contribution
in
Fig. 1 d, f and g
has different color structure from the factorizable terms in Fig. 1a, b, c
and e. The
condensates up to dimension six are kept in our calculation. We also
expand the strange quark
propagator and keep
perturbative term of order ${\cal O}(m_s)$. The calculation is standard
and we simply
present final results after making the double Borel transformation.   

\section{Duality Assumption}
\label{duality}

We may write the dispersion relation for the three-point correlator 
$\Gamma (\omega, \omega')$ as
\begin{equation}
\label{13}
\Pi(\omega, \omega') = \frac{1}{\pi^2}\int_0^{\infty}d\nu\int_0^{\infty}
d\nu'\frac{{\rm Im}\Pi(\nu, \nu')}{(\nu-\omega)(\nu'-\omega')} \; .
\end{equation}
In order to subtract the continuum contribution, we have to invoke quark
hadron
duality assumption and approximate the continuum by the integral over the 
perturbative spectral 
density above a certain energy threshold $\omega_c$.

With the redefinition of the integral variables
\begin{eqnarray}
\label{14}
\nu_+ & = & \displaystyle\frac{\nu+\nu'}{2} \; , \nonumber\\
\nu_- & = & \displaystyle\frac{\nu-\nu'}{2} \; ,
\end{eqnarray}
the integration becomes
\begin{equation}
\label{15}
\int_0^{\infty}d\nu\int_0^{\infty}d\nu'... = 
2\int_0^{\infty}d\nu_+\int_{-\nu_+}^{\nu_+}d\nu_- ... \; .
\end{equation}
It is in $\nu_+$ that the quark-hadron duality is assumed \cite{1,2,3},
\begin{equation}
\label{16}
{\rm higher~~states} = \frac{2}{\pi^2}\int_{\omega_c}^{\infty}d\nu_+
\int_{-\nu_+}^{\nu_+}d\nu_-
\frac{{\rm Im}\Pi(\nu, \nu')}{(\nu-\omega)(\nu'-\omega')} \; .
\end{equation}
This kind of assumption was suggested in calculating the Isgur-Wise 
function in Ref. \cite{2} and was argued for in Ref. \cite{3}.
As pointed out in \cite{1,3}, in calculating three-point functions 
the duality is valid after integrating the spectral density over the 
"off-diagonal" variable $\nu_-=\frac{1}{2}(\nu-\nu')$. Such a duality 
assumption is favored over the naive one:
\begin{equation}
\label{16-1}
{\rm higher~~states} = \frac{1}{\pi^2}\int^{\infty}_{\omega_c} d\nu
\int^{\infty}_{\omega_c}d \nu' 
\frac{{\rm Im}\Pi(\nu, \nu')}{(\nu-\omega)(\nu'-\omega')} \; .
\end{equation}

\section{QCD sum rules}

The spectral density $\rho_{L,S}(s_1, s_2)$ of the perturbative term reads
\be
\rho_L(s_1, s_2)={N_c(N_c+1)\over 2\pi^4}s_1 s_2[s_1s_2 +m_s(s_1+s_2)]
\ee
\be
\rho_S(s_1, s_2)={N_c(2N_c-1)\over 4\pi^4}s_1 s_2[s_1s_2 +m_s(s_1+s_2)]
\ee

The sum rule for $\langle {\bar B}_s| O_{L,S}|B_s\rangle$ 
after the inclusion of the condensates and the integration with the 
variable $\nu_-$ is
\begin{eqnarray}
\label{17}\nonumber
{F_{B_s}^2\over 4}\langle {\bar B}_s|
O_L|B_s\rangle \exp\left(-\frac{\bar{\Lambda}}{T}\right)
 & = & 
{N_c(N_c+1)\over \pi^4}\{\displaystyle \int_0^{\omega_c}d\nu
\exp\left(-\frac{\nu}{T}\right)
 [\frac{16}{15} \nu^5 +{8\over 3}m_s\nu^4]\\ \nonumber
&&+{4\over 3}a_s T^3(1-{m_0^2\over 64 T^2})
+{1\over 6}m_s a_s T^2+{a_s^2\over 288}\}\\
&& -{N_c^2-1\over 256\pi^4}[2T^2\langle g^2_sG^2 \rangle
+a_s m_0^2 T] \;,
\end{eqnarray}
where $a_s =-(2\pi)^2 \langle\bar{s}s\rangle$ and we have used the
factorization
assumption for the four-quark condensates. 
Similarly we have
\begin{eqnarray}
\label{18}\nonumber
{F_{B_s}^2\over 4}\langle {\bar B}_s|
O_S|B_s\rangle \exp\left(-\frac{\bar{\Lambda}}{T}\right)
 & = & 
{N_c(2N_c-1)\over 2\pi^4}\{\displaystyle \int_0^{\omega_c}d\nu
\exp\left(-\frac{\nu}{T}\right)
 [\frac{16}{15} \nu^5 +{8\over 3}m_s\nu^4]\\ \nonumber
&&+{4\over 3}a_s T^3(1-{m_0^2\over 64 T^2})+{1\over 6}m_s a_s
T^2+{a_s^2\over 288}\}\\
&& -{N_c^2-1\over 512\pi^4}[2T^2\langle g^2_sG^2 \rangle
+a_s m_0^2 T] \;.
\end{eqnarray}
We want to emphasize that in Eqs. (\ref{17}), (\ref{18}) the terms with
color factor
$N_c (N_c+1)$ and $N_c(2N_c-1)$ come from the factorizable diagrams in
Fig. 1 a, b, c and e.
The non-factorizable contribution has a color factor ${N_c^2-1\over 2}$
which comes 
from the summation over color factor, $\mbox{Tr} [{\lambda^a\over
2}{\lambda^a\over
2}]={N_c^2-1\over 2}$, in Fig. 1 d, f and g. A second observation is that
the factorizable terms
are all positive while nonfactorizable pieces are negative.

Now we turn to the numerical analysis. The decay constant and binding
energy of the $B_s$ meson at the leading order of heavy quark expansion
can be obtained from the mass sum rule \cite{luo}. 
\begin{equation}\label{19}
F^2_{B_s}\exp\left(-2\frac{\bar{\Lambda}}{M}\right)={3\over 8\pi^2}
\int_0^{s_0}ds s (s+2m_s)e^{-s/M}-<\bar s s>(1-{m_0^2\over 4M^2})
\end{equation}
Note $M=2T, s_0=2\omega_c$.
We have not included $\alpha_s$ corrections in Eq. (\ref{19}), 
because they are also neglected in the sum rule for $\langle {\bar B}_s|
O_{L,S}|B_s\rangle$
 (\ref{17})-(\ref{18}). The values of the parameters are calculated to be
$ F_{B_s} =(0.49\pm 0.1)$ GeV$^{3/2}$, 
$\bar\Lambda =(0.68\pm 0.1) $GeV with 
the threshold $s_0$ to be $(2.2\pm 0.3) $GeV and 
the Borel parameter $M$ in the window ($0.65-1.05$) GeV \cite{luo}.
Numerically we use the following values of the condensates,
\begin{equation}
\begin{array}{lll}
\langle\bar{s}s\rangle&\simeq&-0.8 \times (0.23~ {\rm GeV})^3~,\\
\langle g^2_sG^2\rangle&\simeq&0.48~ {\rm GeV}^4~,\\
\langle g\bar{s}\sigma_{\mu\nu}G^{\mu\nu}s\rangle&\equiv&
m_0^2\langle\bar{s}s\rangle~,
~~~~~m_0^2\simeq0.8~{\rm GeV}^2~.\\[3mm]
\end{array}
\end{equation}
For the strange quark mass we use $m_s=0.15$ GeV.

In order to minimize the dependence of the parameters we 
divide Eqs. (\ref{17}, \ref{18}) by Eq. (\ref{19}) 
to extract
the matrix elements, the variation of which with $\omega_c$ 
and $T$ are given in Fig. 2 and 3.  
The sum rule window is $T=(0.2 - 0.5 )$ GeV, which is almost 
the same as that in the two-point correlator sum rule.
We obtain 
\be
\langle {\bar B}_s|O_L|B_s\rangle=(0.85\pm 0.20) \mbox{GeV}^4 \; , 
\ee
\be
|\langle {\bar B}_s|O_S|B_s\rangle|=(0.55\pm 0.15) \mbox{GeV}^4 
\ee
where the central value corresponds to $T=0.3$GeV and
$\omega_c=1.1$GeV. The uncertainty 
includes the variation with $T$ and $\omega_c$. 
The bag parameters
B and $B_S$ are defined by 
\be
\langle {\bar B}_s|O_L|B_s\rangle=\frac{8}{3}f^2_{B_s}M^2_{B_s}B,\; \; 
\langle {\bar B}_s|O_S|B_s\rangle=- \frac{5}{3}f^2_{B_s}M^2_{B_s}
\frac{M^2_{B_s}}{(\bar{m}_b+ \bar{m}_s)^2}B_S.
\ee
and they can be directly obtained from Eqs. (22), (23) and (24).

The ratio of these two matrix elements is very interesting. We divide
Eq. (19) by  
Eq. (18) to extract the numerical value of the ratio. In such a way the
dependence on the 
the Borel parameter and the continuum threshold is minimized as can be
clearly
seen in Fig. 4. Within the accuracy of QSR the curve in Fig. 4 is
flat. The ratio
is practically the same in the working region of $T$ and $\omega_c$. It
reads
\be\label{35}
{\cal R} ={|\langle {\bar B}_s|O_S|B_s\rangle|\over \langle {\bar
B}_s|O_L|B_s\rangle}=(0.63\pm 0.13)
\ee

In our numerical calculation, the contribution of the perturbative term is
about $45-65\%$ of the total contributions in preferred Borel variable
region. We have used factorization approximation for the four quark condensates in
numerical calculations. This may
introduce some uncertainty. We may introduce a scale factor $\kappa$ to
indicate the 
deviation from the factorization approximation as in \cite{zhu}. 
In our calculations of sum rules the $1/M_b$ corrections in HQET have
not been included in
, which may bring a deviation from the numerical results of the 
matrix elements. However, for the ratio of the two matrix elements, we
expect little change to the above analysis. Our numerical results are not
sensitive to the mass of the strange quark. Actually, the effects due to
the strange quark are very small so that the results for $B_s$ are almost
the same as those for $B_d$.

We now give a remark on the usual factorization assumption. In
our Feynman diagram  (Fig. 1) calculations, the 
contributions of nonfactorizable diagrams
are around $-6\%, -7\%$ for $\langle {\bar
B}_s|O_S|B_s\rangle$ and $\langle {\bar B}_s|O_L|B_s\rangle$ respectively,
which means that the factorization approach works well even
though our calculations are limited to the leading order in the 1/$m_b$ expansion in HQET.
That is, the conclusion in Ref.\cite{narison} remains unchanged when the
nonperturbative 
condensate contributions are taken into account. If one considers
$\alpha_s$ corrections, there is only one nonfactorizable perturbative
diagram, in which the gluon line in Fig. 1 f is
connected, in the fixed-point gauge in the leading of $1/M_b$ expansion.
However, radiative corrections are generally of high order ${\alpha_s 
\over \pi}$ compared to the leading order, and the fact is that the 
perturbative term is about
$45-65\%$ to the whole contribution, so the contribution from the diagram
can be neglected compared to those in Fig. 1 (d), (f) and (g). The case
here is different with that in the
calculation of matrix elements of four-quark operators, relevant to the
life time difference between heavy mesons, where the flavor changes
$\Delta F=0$. In that case, the perturbative contribution
vanishes\cite{liu}, and we
can't predict naively how large the radiative correction is compared to the
nonperturbative terms. 

\section{The $B_s$ and ${\bar B}_s$ decay width difference}

The complete expression for $\Delta\Gamma_{B_s}$ with 
short-distance coefficients at NLO in QCD is given by \cite{ben}
\begin{eqnarray}\label{dgg}
&&\left(\frac{\Delta\Gamma}{\Gamma}\right)_{B_s}=
\frac{16\pi^2 B(B_s\to Xe\nu)}{g(z)\tilde\eta_{QCD}}
\frac{f^2_{B_s}M_{B_s}}{m^3_b}|V_{cs}|^2\cdot \\
&&\cdot\left(G(z)\,\frac{8}{3}\,B+
    G_S(z)\,\frac{M^2_{B_s}}{(\bar m_b+\bar m_s)^2}\,\frac{5}{3}
    \,B_S
   +\sqrt{1-4z} ~\delta_{1/m}\right), \nonumber
\end{eqnarray}
where
\begin{equation}\label{ggs}
G(z)=F(z)+P(z)\quad \mbox{and}\quad G_S(z)=-(F_S(z)+P_S(z)).
\end{equation}
and F, P, $F_S$, $P_S$ can be found in Ref. \cite{ben}.
We eliminated the total decay rate $\Gamma_{B_s}$
in favor of the semileptonic branching ratio $ B(B_s\to Xe\nu)$, 
as done in \cite{beneke}.
This cancels the dependence of $(\Delta\Gamma/\Gamma)$ on $V_{cb}$
and introduces the phase space function
\begin{equation}\label{gz}
g(z)=1-8z+8z^3-z^4-12z^2\ln z,
\end{equation}
as well as the QCD correction factor \cite{CM}
\begin{equation}\label{etqcd}
\tilde\eta_{QCD}=1-\frac{2\alpha_s(m_b)}{3\pi}
\left[\left(\pi^2-\frac{31}{4}\right)(1-\sqrt{z})^2+\frac{3}{2}\right].
\end{equation}
One can also express the width difference as
\bea
\label{eq:present} 
{\Delta \Gamma_ {B_s}\over  \Gamma_ {B_s}} \ = \ \left( \tau_{B_s}  
\Delta m_{B_d} \frac{m_{B_s}}{m_{B_d}} \right)^{\rm (exp.)}\ 
\left| V_{ts} \over V_{td} \right|^2 {K} \cdot \biggl( G(z) - G_S(z)
{\cal R} (m_b) + \tilde \delta_{1/m} \biggr)
\ \xi^2 \ ,
\eea
where 
\bea \xi= \frac{ f_{B_s}
\sqrt{\hat B_{B_s}}}{f_{B_d}\sqrt{\hat B_{B_d}}} \ , \eea 
$K$ is the Eq. (7) in ref. \cite{lattice},
\bea \tilde \delta_{1/m} = f^2_{B_s}M^2_{B_s}\delta _{1/m}, \eea
and the $\tilde \delta_{1/m}$ represents the 1/$m_b$ corrections and can
be found in
Ref. \cite{beneke}.

It is clear from the above equation that  besides the ratio ${\cal R}$  of
the matrix 
elements of four quark operators, 
 which are those we have calculated in the paper,
we only use the experimental $B_d$-meson mass difference, which is known
with a tiny error 
\cite{schneider}
\bea
 \left( \Delta m_{B_d} \right)^{\rm (exp.)} = 0.484(15)\, ps^{-1} \ ,\eea
and another ratio of hadronic matrix elements, $\xi$, which is
rather accurately determined in lattice simulations
\cite{reviews,ourBBar}. 

As it is well known, the quantities in Eq. \ref{dgabc} are
calculated at the scale O($m_b$), while our result Eq. (\ref{35}) is
calculated at  the hadron scale
$\mu_{had}$. Therefore, We have to consider the renormalization scale
dependence of those four-quark operators. The anomalous dimension matrix of these operators
 has been given in Ref. \cite{ben}.  Using the anomalous dimension matrix and following the
standard way, we 
 obtain the scale dependence of ${\cal R}$  
\be
{\cal R}(m_b)=1.69{\cal R}(\mu_{had})+0.03.
\label{sca}
\ee
where ${\cal R}(\mu_{had})$ is defined by Eq. (\ref{35}). To obtain the
numerical result, $m_b=4.8$
GeV and $\mu_{had}=1.0$ GeV have been used. It is obvious from
Eq. (\ref{sca}) that the result depends 
on the renormalization scale heavily.

Numerically we have 
\bea \nonumber
{\Delta \Gamma_ {B_s}\over  \Gamma_ {B_s}}&=[(0.5\pm 0.1)+(13.8\pm
2.8) {\cal R}(m_b)+(15.7\pm 2.8)
(-0.55\pm 0.17)]\times 10^{-2}\\
&=(7.0\pm 0.8)\times 10^{-2}\label{num}
\eea
Clearly such a life time difference is compatible with existed
literatures.
It's interesting to compare our result to the two recent
lattice QCD calculation: 
${\Delta\Gamma_ {B_s}\over\Gamma_{B_s}}=(10.7\pm 2.6\pm 1.4\pm 1.7)\times
10^{-2}$
in Ref. \cite{japan} and
${\Delta \Gamma_ {B_s}\over \Gamma_{B_s}}=(4.7\pm 1.5\pm 1.6)\times
10^{-2}$ 
in Ref. \cite{lattice}. In Eq. (\ref{num}) the numerical value of
$\tilde{\delta}_{1/m}$, 
which corresponds to the $1/m_b$ correction in the short distance
expansion of the operator product $H_{eff}(x)H_{eff}(0)$
\cite{ben}, has been taken as -0.55\cite{lattice}. If it is taken as
-0.30, one has $\Delta\Gamma/\Gamma=10.9 \times 10^{-2}$, larger than
$7.0 \times 10^{-2}$, while in the case of
 Ref. \cite{lattice}, $\Delta\Gamma/\Gamma$ would remain in the $10\%$
range with the change from -0.55 to -0.30. That is,
the sensitivity to the final term in Eq. (\ref{num}), i.e., the $1/m_b$
correction, increases in our result. Without a good control of
the correction, a precise determination of the lifetime difference is 
impossible.

\section{Conclusion and discussion}
In summary we have calculated the matrix elements of the four-quark
operators relevant to the
$B_s$ meson lifetime difference in QCD sum rules in HQET. The
sum rules are constructed with three-point correlators  and both the
perturbative
and nonperturbative contribution are taken into account. Our result
shows that the usual factorization assumption is indeed a good
approximation.
The numerical results show that the sum rules of those operators have a
good platform. The perturbative contribution to sum rules are about $45-65\%$
of the total  contribution. Our results are not sensitive to $m_s$. 
The life difference ${\Delta\Gamma_{B_s}\over \Gamma_{B_s}}$ is
found to be around $(7.0\pm 0.8)\times 10^{-2}$. This result is compatible
with those predicted by lattice calculations. 
The $\alpha_s$ corrections have not been taken into account in the sum rules
and they will definitely have effects on the
resulting numerical values. To get more accurate prediction, the $\alpha_s$
corrections should be taken into account, which is beyond the content of
the letter.

\section*{Acknowledgments}
The work was partly supported by the National Natural Science
Foundation of China.

\newpage
\begin{figure}[htb]
\begin{center}
\epsfig{file=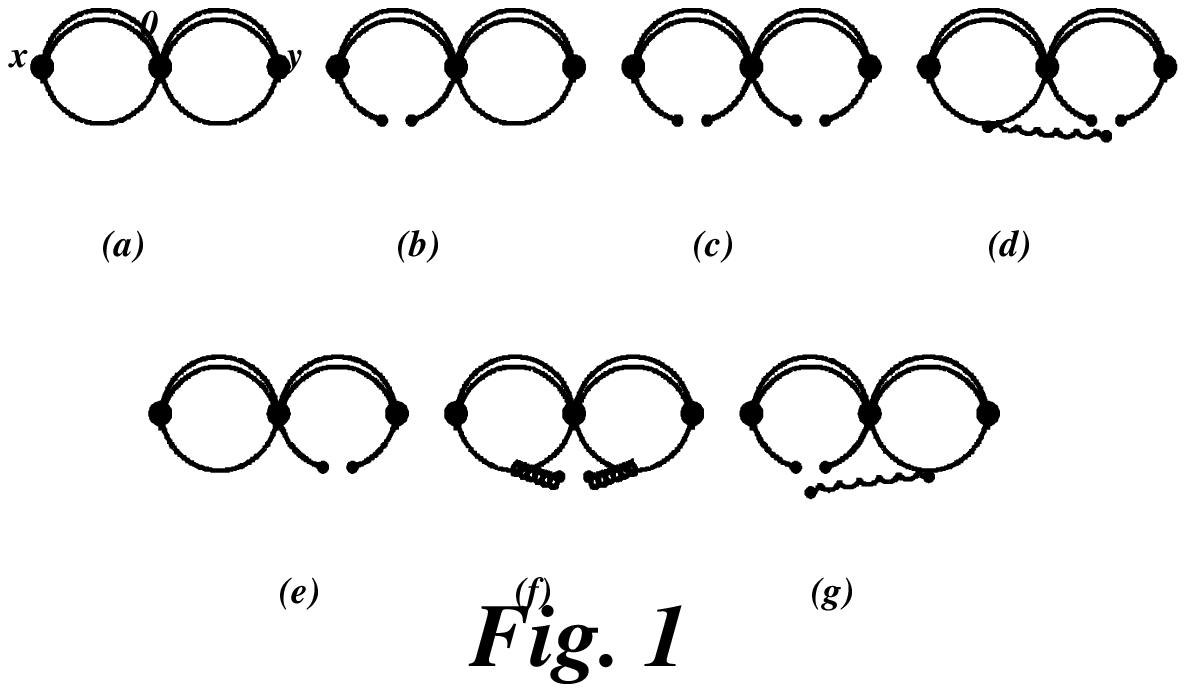,width=8cm}
\end{center}
\caption{ Dominant non-vanishing Feynman diagrams for
$\Gamma^O(\omega, \omega')$
}
\end{figure}

\vspace{1.2 cm}

\begin{figure}[htb]
\begin{center}
\epsfig{file=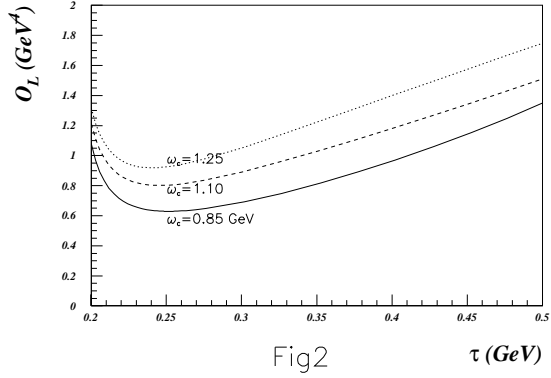,width=8cm}
\end{center}
\caption{ The dependence of $\langle {\bar B}_s|O_L|B_s\rangle$ on $T,
\omega_c$
}
\end{figure}

\vspace{1.2 cm}

\begin{figure}[htb]
\begin{center}
\epsfig{file=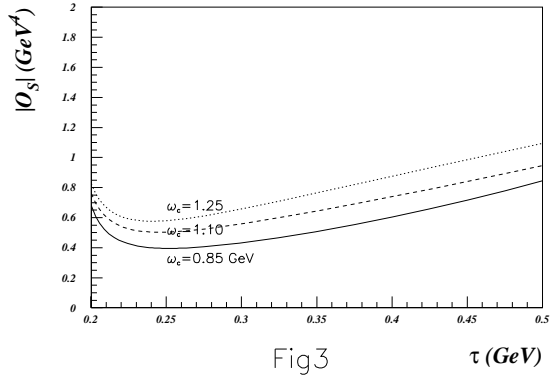,width=8cm}
\end{center}
\caption{ The variation of $|\langle {\bar B}_s|O_S|B_s\rangle|$ with $T,
\omega_c$.
}
\end{figure}  

\vspace{1.2 cm}

\begin{figure}[htb]
\begin{center}
\epsfig{file=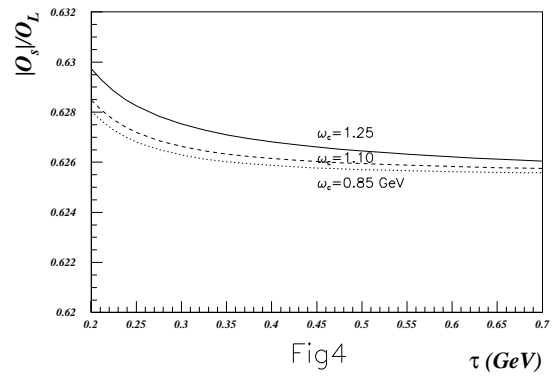,width=8cm}
\end{center}
\caption{ The variation of ${\cal R}$ with $T, \omega_c$
}
\end{figure}

\end{document}